# Colloidal Model for Investigating Optimal Efficiency in Weakly Coupled Ratchet Motors

José Martín-Roca
*Departamento de Estructura de la Materia, Física Térmica y Electrónica, Universidad Complutense de Madrid, 28040 Madrid, Spain.*

Laura Izquierdo Solis and Fernando Martínez Pedrero*
*Departamento de Química-Física, Universidad Complutense de Madrid, Avda. Complutense s/n, 28040 Madrid, Spain.*

Pau Casadejust, Ignacio Pagonabarraga, and Carles Calero†
*Departament de Física de la Matèria Condensada, Facultat de Física, Universitat de Barcelona, Barcelona, 08028 Spain.*

We investigate the transport of superparamagnetic colloidal particles along self-assembled tracks using a periodically applied magnetic field as a model for ratchet-like mechanisms. Through video microscopy and simulations, we examine how factors such as particle size, track roughness, and the shape, intensity, and frequency of the applied potential influence transport efficiency. The findings reveal that processive motion can be achieved without residual attraction, with optimal transport efficiency governed by the combined effects of particle size ratios, actuation frequency, track roughness, and asymmetry in the applied potential. Additionally, we explore alternative strategies, including weak residual attraction and alternating magnetic fields, to further enhance efficiency. These findings provide valuable insights for the development of synthetic micro/nanomotors with potential applications in drug delivery and environmental remediation.

Biological motors are enzymes that convert chemical and thermal energy into directed mechanical motion through ratchet-like mechanisms. These enzymes actively interact with cytoskeletal filaments, which are polarized structures extending over micrometer distances. The movement involves alternating between two states: a detached state where the enzymes diffuse, and another where they engage the filaments under an asymmetric potential. The asymmetry of the potential ensures that, upon reactivation, the distance required to reach the next binding site in one direction is shorter than in the opposite direction. This process rectifies Brownian motion, enabling the efficient transport of cargo, muscle fibers, viruses or chromosomes. These motors are fueled by the binding and hydrolysis of adenosine triphosphate (ATP), releasing approximately 22 $k_BT$ of energy. This energy facilitates the cyclic binding and unbinding of the motor to the biopolymer, driving unidirectional movement at speeds around 1 $\mu m/s$ [1].

The dimeric structure of motors with two heads enhances their processivity, allowing them to take multiple steps along the filament before detachment [2]. These motors achieve processive movement through a 'hand-over-hand' mechanism, where the heads alternate their leading position with each step [3]. In contrast, monomeric motors typically lack processivity due to their low duty ratios and spend a significant portion of their cycle detached. However, exceptions exist and studies on KIF1A, a monomeric motor, reveal that single-headed motors can exhibit processivity similar to double-headed ones, displaying bidirectional stepping with a bias toward forward movement, consistent with the noise-induced transport of a flashing ratchet model [4, 5].

Despite extensive research, many aspects of motor dynamics remain unresolved due to the small size, diversity, and complexity of these systems [7]. Several uncertainties remain to be addressed, including the precise details of protein structures, the mechanisms by which ATP-derived energy is converted into mechanical work, and the influence of various factors such as thermal fluctuations, irregularities in the track, and the ATP hydrolysis rate, to name a few. All these factors determine the efficiency of transport in molecular motors [8–10] and understanding their effect is crucial for unlocking the full potential of molecular-scale transport systems, which could lead to groundbreaking and efficient solutions in this rapidly advancing field [11].

Colloidal model systems offer a powerful means to attain a deeper comprehension of such complex biological phenomena, which are hard to observe experimentally. Due to their size, colloidal particles can be directly observed with optical microscopy, as well as indirectly by light scattering. In addition, their tunability and the possibility of external control [12] makes them ideal systems to study a range of phenomena in condensed matter physics, including the glass transition [13], crystal nucleation [14], or the frustration in ice manifolds [15] to mention just a few examples. Conversely, the motion strategies of biological systems have inspired advancements in colloidal science and nanotechnology [16–18]. One notable application is the creation of a flashing ratchet potential using periodically asymmetric geometric patterns etched onto surfaces and combined with transverse electric fields [19], or the enhancement of motility in magnetic microparticles through asymmetric actuation timescales [20, 21]. Similar principles are employed in surface-mounted molecular rotors, which are driven by alternating electric fields, light pulses or rocking mechanism [22, 23]. The possibility of net transport of colloidal systems under flashing ratchet potentials was demonstrated experimentally, with time-dependent electric potentials [24] and optical traps modulated to generate sawtooth potentials [25], and promises for innovative biomedicine applications [1, 26].

* fernandm@ucm.es
† carles.calero@ub.edu

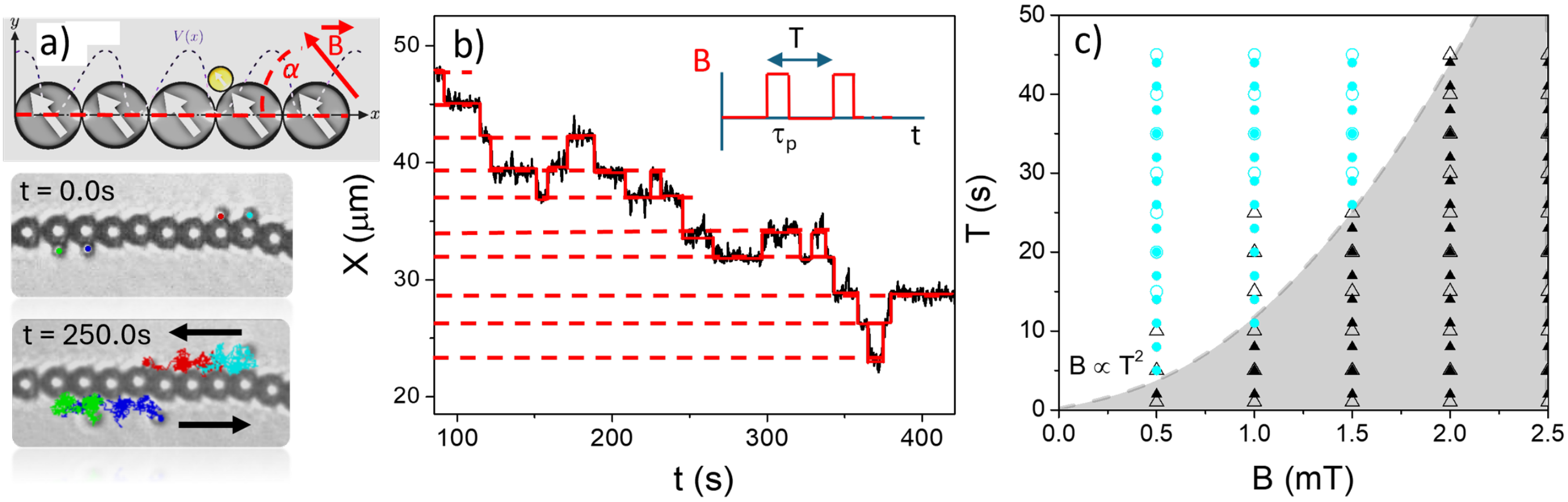

FIG. 1. a) Superparamagnetic colloidal particles (radius $a' = 0.5\ \mu m$), are transported along self-assembled linear tracks of larger particles (radius $a = 1.4\ \mu m$), by applying periodic magnetic field pulses tilted at an angle $\alpha$ relative to the chain axis. The particles preferentially move along the chain in the direction favored by the field's inclination. b) The displacement of the moving particle occurs in discrete steps of size $2a$, mostly directed towards decreasing values of $X$ as favored by the asymmetry of the potential, although multiple occurrences of steps in the opposite direction are also observed. c) The escape phase diagram for $a = 1.4\ \mu m$, $a' = 0.5\ \mu m$, $\alpha = 90^o$, is constructed by varying magnetic field strengths and periods, with a fixed pulse duration $\tau_p = T/3$. In experiments (empty points), particles are deemed to process (black triangles) or escape (blue circumferences) if they fail to return to the potential minima generated by the hosting chain in at least one of the first 20 periods. Each filled point represents the average from $N = 600$ simulations, each running for $t = 20T$. Here, escape points (blue filled circles) indicate particles escaped in 90% or more of simulations. The dashed line represents the $T \propto B^2$ dependence detailed in the Supplementary Materials [6].

In this study, we investigate a magnetic colloidal system designed as a model to study ratchet transport of molecular motors on filaments. The system consists of mobile paramagnetic spherical colloidal particles in contact with a self-assembled linear chain of paramagnetic particles pre-adsorbed onto a solid substrate along the X-axis, see Fig. 1a. The colloidal particles employed are superparamagnetic M270-Dynabeads and MyOne, Invitrogen with radii 1.4 and 0.5 $\mu m$, respectively (see Supplementary Materials [6] for more details). Due to sedimentation, the mobile colloids predominantly remain in contact with the solid substrate and their dynamics is nearly two-dimensional. In the absence of an external magnetic field, they exhibit essentially diffusive dynamics, only altered by excluded volume interactions with the fixed chains of particles. In the presence of a magnetic field, the dynamics of mobile colloids is determined by the spatially periodic dipolar interaction with the adsorbed fixed chain, whose potential exhibits an asymmetric profile if the field is in the plane of the substrate and forms an angle $\alpha \in (0, \pi/2)$ with the X-axis (see Fig. 1a). The alternation of such external actuation, over a period $\tau_p$, with periods without field, drives the transport of mobile colloids along discrete trajectories which resemble those of weakly coupled molecular walkers. In Fig. 1b we show the trajectory of a single mobile colloid (radius $a' = 0.5$ $\mu$m) along a fixed chain of superparamagnetic particles (radius $a = 1.4\ \mu m$) as periods $\tau_p = T/3$ of external actuation with a field forming an angle $\alpha = \pi/4$ are alternated with periods $2T/3$ with no actuation. The displacement of the moving particle occurs in discrete steps of size $2a$, mostly directed towards decreasing values of $X$ as favored by the asymmetry of the potential, although occasional steps in the opposite direction are also observed. The dynamics in steps of the moving colloids along the magnetic filament exhibits the same basic features as the motion of molecular motors on microtubules, although the size of the steps differ significantly in scale. Molecular motors such as kinesin typically take steps of $\approx$ 8nm, as observed both in vitro and in vivo [4, 28–30]), while the colloidal system performs steps nearly three orders of magnitude larger. This size disparity greatly facilitates direct observation and analysis of the dynamics in the colloidal model. Video microscopy enables tracking particle trajectories and analyzing factors affecting transport efficiency, such as the strength, angle, and the period of application of the external field, or structural characteristics like the degree of linearity of the filament, the sizes of the moving magnetic particles, and of those forming the fixed magnetic filament. Experimental results were systematically compared with Brownian dynamics simulations, which accurately reproduced the characteristic trajectories [6]. This controlled model system offers insights into phenomena that are otherwise challenging to explore in biological motors.

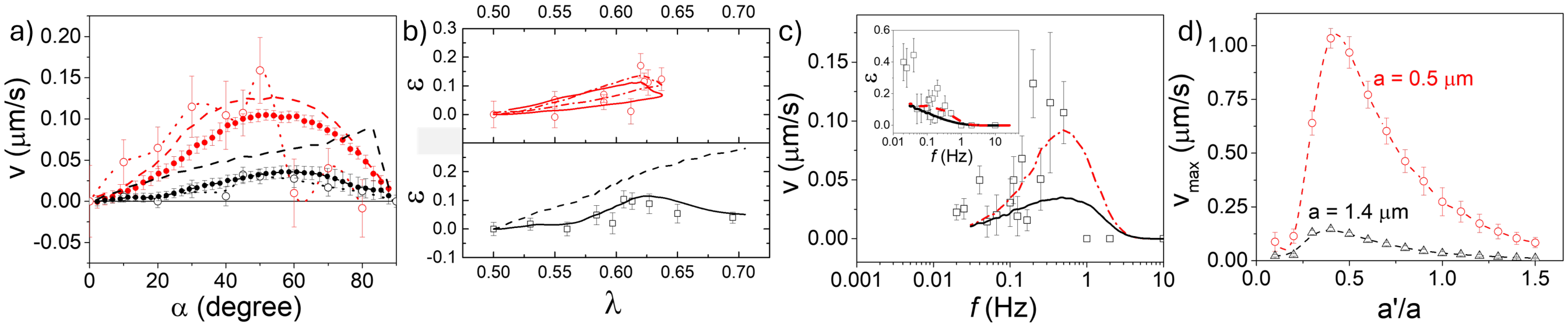

FIG. 2. a) The dependence of the average velocity $v$ on $\alpha$ is shown for experiments (discrete points) and simulations on both linear (dashed lines) and irregular track geometries (dotted lines). The irregularity included in the simulations, characterized by $\sigma = a/3.5$, matches experimental measurements. b) Mechanism efficiency $\epsilon = vT/2a$, against $\lambda$, a dimensionless parameter describing the asymmetry of the potential generated by a straight chain. $\lambda = 0.5$ denotes a symmetrical potential, while $\lambda = 1$ represents a monotonously varying periodic potential (Supplementary Materials [6] and reference [27]). In a) and b) all simulations and experiments used $a' = 0.5$ $\mu m$ (red) or $a' = 1.4$ $\mu m$ (black), with $B = 2.0$ $mT$ and field periods of 3.0 s for 0.5 $\mu m$ particles and 9.0 s for 1.4 $\mu m$ particles, with pulse duration of $\tau_p = T/3$. In b), the arrows indicate the direction of increasing $\alpha$. c) $v$ vs pulse frequency, $f$, for particles with $a' = 1.4$ $\mu m$ and $\alpha = 45^o$, $B = 2.0$ $mT$, and $\tau_p = 1/(3f)$, in experiments (discrete points), and simulations conducted both on linear (red dash-dotted lines) or irregular track geometries (black continuous lines). The irregularity is again characterized by $\sigma = a/3.5$, similar to that measured in the experiments. The inset shows the corresponding dependence of $\epsilon$ on $f$. d) Maximum velocity measured, $v_{max}$, by changing the pulse frequency, with respect to the size ratio between the mobile particles and chain particles. The parameters used are $\tau_p = T/3$, $\alpha = 45^o$, $B = 2.0$ mT for $a = 1.4$ $\mu m$ and $B = 9.4$ mT for $a = 0.5$ $\mu m$. Each point is the average of simulations done in 10 different chain configurations. For each configuration, we do 300 independent simulations, where each simulation runs for $15T$.

The process of detachment of molecular motors and its effect on the ability of transport along microtubules is still unclear [31]. The ability to perform processive motion ranges across different molecular motors with different molecular structures, typically losing processivity in their monomeric form [4]. While the processivity of molecular motors is mainly determined by their molecular structure, the colloidal model system enables precise external control over the detachment dynamics of the magnetic colloid, facilitating its movement via ratchet transport. A measure of detachment from the fixed linear chain, which serves to characterize the processivity of the colloidal model, is provided in Fig. 1c, as a function of magnetic field strength $B$ and period of actuation $T$. In this diagram we can identify a range of parameters for which the mobile colloid eventually detaches from the magnetic filament. Given the magnitude and orientation of the applied magnetic field, moving colloids can escape the track if after the diffusive period their thermal energy exceeds the magnetic attraction energy as the field is activated. This leads to a critical pulse period $T_c \sim B^2$ [6] which distinguishes actuations with periods $T < T_c$, for which the mobile colloids exhibit long processive motion (remain attached to the filament for long periods of time), and periods $T > T_c$, for which the mobile colloids detach rapidly.

The nature of the interaction between the motor and the filament can also be tuned in the colloidal model system. The asymmetry of the potential acting on the mobile magnetic colloid during the actuation period depends on the angle $\alpha$ formed by the external field with the axis of the magnetic filament, which determines the velocity of transport. In Fig. 2a we examine the dependence of the averaged velocity $v$ (averaged over at least 100 pulses) on $\alpha$ using fields and periods of actuation for which the mobile colloid does not escape the filament. Error bars represent the standard deviation of a trinomial distribution, as particles typically remain stationary, recede, or advance by one particle in the adsorbed chain [6]. Experimental measurements are compared with results from simulations conducted on both perfectly straight and irregular jagged tracks. The irregularity of the track is quantified by the standard deviation $\sigma$ of the distance between the centers of the particles and the main axis of the chain, with $\sigma = a/3.5$ matching the experimental measurements. Fig. 2a shows that 0.5 $\mu m$ particles reach the maximum velocity at $\alpha \sim 50^o$. For $\alpha < 50^o$, the experimental data align with the simulation trends, regardless of the irregularity of the track. For $\alpha > 50^o$, deviations are attributed to statistical variations and the specific geometry of the chains involved, with staggered intervals that particularly affect larger particles. In fact, it is important to note that particles transported on different chains with identical $\sigma$ values may exhibit varying $v$ vs.$\alpha$ trends [6]. The experimental trend of the 1.4 $\mu m$ particles aligns with simulations on irregular tracks, while simulations on straight tracks show a sharp maximum at $\alpha \sim 80^o$, consistent with theoretical predictions [6]. As $\alpha$ approaches 0 or $90^o$, the velocity drops to zero due to the loss of spatial asymmetry.

The efficiency of colloidal transport, defined as the ratio of the average velocity to the velocity of a walker that advances one step on the chain every period $T$, $\epsilon = vT/2a$, depends on the angle $\alpha$ and thus the asymmetry of the periodic potential. Fig. 2b plots the efficiency against $\lambda$, a dimensionless parameter describing potential asymmetry, where $\lambda$ ranges from 0.5, symmetric potential, to 1, when the periodic potential varies monotonously along the chain between two particles of the track [6, 27]. For $B = 2$ mT and 0.5 $\mu$m particles, simulations match experiments, showing higher $\epsilon$ with greater $\alpha$ and peaking at $\lambda \sim 0.625$. For 1.4 $\mu$m particles, simulation $\epsilon$ aligns with experimental results on rough tracks, peaking at $\lambda \sim 0.625$, while $\epsilon$ on linear tracks increases monotonically with $\lambda$ in the explored $\lambda$ range. Similar $\epsilon$ values are observed for transported particles of different sizes.

In molecular motors, factors such as the ATP hydrolysis rate, filament affinity, load carried by the motor, subunit coordination, temperature, and the spatial periodicity of the ratcheted potential collectively define the timescale of actuation required to achieve optimal transport [32–34]. In the colloidal model system, the period of the actuation $T$ can be independently adjusted. In Fig. 2c we show the dependence of the velocity of transport $v$ on the frequency of actuation, $f = 1/T$, for particles with $a' = 1.4$ $\mu m$ and $\alpha = 45^o$. At high frequencies, velocity drops to zero as particles lack sufficient time to reach the nearest energy minimum within a single cycle. As frequency decreases, more particles reach the adjacent minima, increasing their velocity, with a peak speed around $f \sim 0.5$ Hz. At lower frequencies, and $\tau_p$ constant, particles spend more time attached to the chains, reducing their speed. However, as the frequency decreases further, particles have more chances to explore distant positions on the chain, enhancing the mechanism's effectiveness (see inset in Fig. 2a). At very low frequencies, $f < 0.01$ Hz, mobile particles often escape the chain's influence. In this case, simulations that ignore chain irregularities align better with experimental data, likely due to limited experimental statistics and specific chain geometries. For smaller particles, the curve shifts to higher frequencies, and they attain higher velocities because the potential they experience is more asymmetric compared to that of larger particles under the given conditions [6].

Figure 2d illustrates the maximum velocity, $v_{max}$, obtained in simulations by varying the pulse frequency as a function of the size ratio between the mobile particles and the chain particles. For an adsorbed chain composed of particles of a given size, the maximum velocity is observed when the size ratio between the transported particle and the adsorbed particles is approximately 0.4, consistent across both chain sizes studied. However, when the particles forming the chains are smaller, this characteristic curve becomes difficult to establish. In these cases, at low $a'/a$ values, the mobile particles are too small to be confined by gravity on the substrate plane, leading them to jump erratically from one side of the chains to the other, which disrupts the transport mechanism.

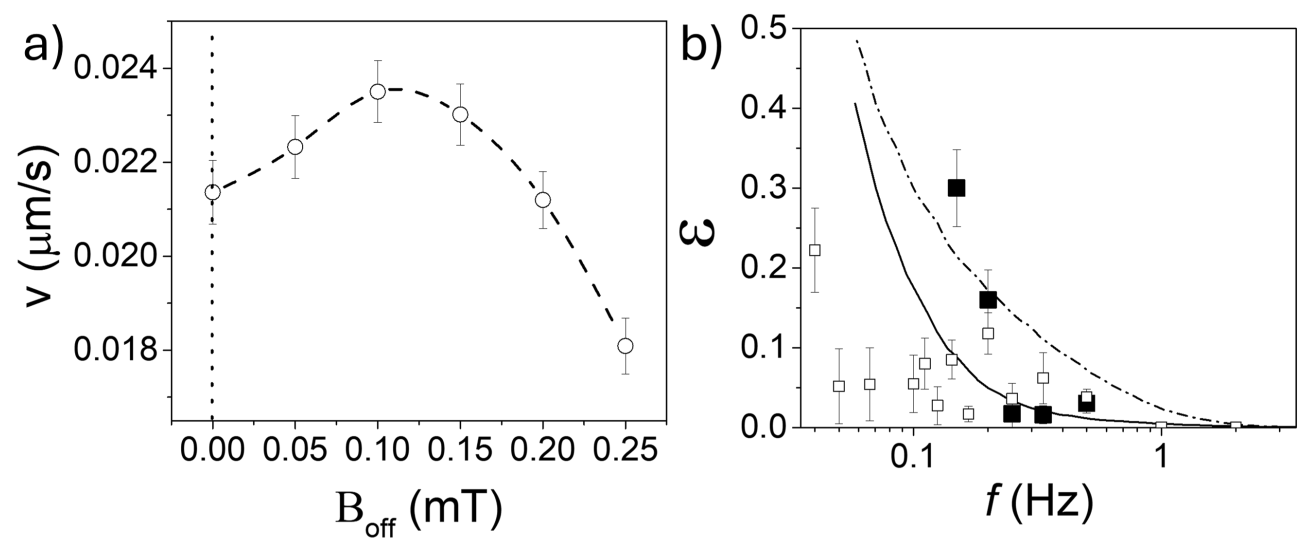

FIG. 3. a) Velocity with respect to the strength of the off magnetic field, $B_{off}$. The parameters used are $a' = a = 1.4$ $\mu m$, $B = 0.5$ mT, $\alpha = 45^o$ for $B$ and $\alpha = 90^o$ for $B_{off}$, $T$= 7 s and $\tau_p = T/3$. The lines are guides for the eye. b) $\epsilon$ vs. $f$ for non-adsorbed particles transported by an alternative ratchet mechanism. In each cycle, a field parallel to the substrate and tilted relative to the chain is applied for $t_{attrac} = T/2$, followed by $t_{rep} = T/2$, during which a field perpendicular to the chain and tilted relative to the confinement plane is applied. The experiments (big squares), are compared with the values measured following the protocol described in Figure 2c (small squares), and with simulations performed on straight chains (continuous line) and roughness chains (dashed lines). $a = a' = 1.4$ $\mu m$, $B_{attrac} = 3.7$ mT, $\alpha = 45^o$, $B_{rep} = 1.7$ mT (pulses for $T/2$).

Some authors demonstrated that different mechanisms and structures emerge from interactions between various KIF1A motors, achieving significant collective effects [5, 35–37]. In the Supplementary Materials [6], we have also shown that the presence of other mobile magnetic colloids can affect the efficiency of the ratchet mechanisms while forming structures compatible with the applied potential. However, it is important to note that the influence of other motile particles in our model system does not directly compare to biological systems. In our colloidal model, all processive motors exhibit synchronized dynamics, attaching to and detaching from the chain simultaneously. Additionally, the dipolar interactions between particles in the system induce the formation of transient linear motile aggregates. These structures assemble during the actuation phase and disassemble during each cycle's diffusive period, further differentiating the behavior of our system from that of molecular motors in biological contexts.

Finally, we explore alternative ratchet strategies to enhance transport efficiency at the nano/microscale. First, we examine using numerical simulations the impact of weak attraction between diffusing particles and the chain during the diffusion period. These simulations are inspired by the KIF1A motor protein's ratchet-like mechanism, which uses electrostatic interactions to create a weakly bound state, enabling one-dimensional Brownian diffusion along a microtubule without detachment [38].

In the colloidal model, our simulations show that the inclusion of a weak residual field $B_{off}$ at a $90^o$ orientation during the diffusive period may enhance the velocity of transport for certain parameters of external actuations, see Fig. 3a). The application of a weak $B_{off}$ induces a weak attraction between the mobile colloid and the filament during the diffusive period, allowing free diffusion in the X-direction while preventing escape events in the Y-direction which are significant for actuations with weak magnetic fields or long periods $T$. Specifically, the application of a residual magnetic field of approximately 0.1 mT increases the average velocity of the motors by 11% compared to the case without the field.

We also explored an alternative approach aiming to reduce diffusion time and improve ratchet efficiency, based on the field induced separation between particles and the chain. In this method, the diffusion period is replaced by a repulsive time interval $t_{rep}$, during which we apply a field perpendicular to the chain, forming an angle $\beta = 95^o$ with the confinement plane. The magnetic torque generated by the reorientation of the field pushes non-adsorbed particles connected to the chains downwards, away from the horizontal plane. However, the presence of the glass substrate hinders this downward displacement. As a result, motile particles remain on the glass surface with their dipoles parallel to those in the chains after the field reorientation. The mobile colloids are initially ejected and, once beyond the interaction range of the fixed chain, where magnetic repulsion becomes comparable to thermal energy, they begin diffusing on the glass surface [39]. As shown in Fig. 3b, this alternative strategy enhances efficiency over the method described in Figure 2c, despite shorter diffusion times at the same frequency. However, in experiments, the cycle frequency must remain between 0.1 and 0.5 Hz, as lower frequencies increase the risk of particle detachment from the track. Simulations show a decrease in $\epsilon$ as the frequency increases, regardless of whether the non-adsorbed particles are transported along straight or irregular chains. Interestingly, in the simulations the mechanism remains processive at lower frequencies. The observed difference could be due to the effect of drift motion due to residual fluid currents that can occur in the experiments, especially noticeable at low frequencies, when the diffusion periods are long.

In conclusion, this study explores the transport of superparamagnetic colloidal particles along self-assembled linear tracks of adsorbed particles under a periodically applied magnetic field. The system serves as an in vitro model to investigate ratchet-like transport mechanisms analogous to those used by biological motors, such as KIF1A, and is controlled by adjustable parameters. In our experiments, magnetic microparticles are exposed to a magnetic field tilted relative to the tracks, creating an asymmetric periodic potential that drives the particles along discrete paths. Video microscopy reveals how factors such as field direction, strength and frequency, track roughness, and particle size affect transport efficiency and processivity, with simulations validating experimental results. Our findings indicate that processive motion can occur without residual attraction during diffusion, with the critical pulse period scaling as $T_c \propto B^2$. In the range of mobile particle sizes studied, the optimum transport efficiency is reached at field frequencies between 0.1 and 0.5 Hz. This efficiency occurs when the size ratio between the transported particles and the particles forming the chain is about 0.4, and when the applied field generates a potential around the chains with an asymmetry parameter of about $\lambda \sim 0.65$. We also demonstrate that, similar to the nanoscale, irregularities in the adsorbed structures and slight particle polydispersity can affect the mechanism's efficiency [40]. In addition, we explore alternative ratcheting mechanisms by introducing a weak residual attraction during diffusion, inspired by motor-track interactions, and test a strategy involving alternating magnetic fields to improve transport efficiency by promoting separation of particles from chains. Understanding this simple colloidal system provides insights into the ratchet transport mechanisms. The knowledge gained can shed light on the complex machinery of molecular motors and inform the design of synthetic micro/nanomotors for biomedical applications [41, 42].

## ACKNOWLEDGMENTS

We thank L. Sánchez-Viso for initial experiments. This work was funded by Ministerio de Ciencia e Innovacion (Grant No. PID2022-140407NB-C21) funded by MCIN/AEI/10.13039/501100011033 and "ERDF A way of making Europe". CC acknowledges support from the Spanish grant PID2021-124297NB-C31 funded by MCIN/AEI/10.13039/501100011033 and "ERDF A way of making Europe", and from the Spanish grant CNS2022-135395 funded by MCIN/AEI/10.13039/501100011033 and by the European Union NextGenerationEU/PRTR. I.P. acknowledges support from Ministerio de Ciencia, Innovación y Universidades MCIU/AEI/FEDER for financial support under grant agreement PID2021-126570NB-100 AEI/FEDER-EU, and Generalitat de Catalunya for financial support under Program Icrea Acadèmia

[1] P. M. Hoffmann, How molecular motors extract order from chaos (a key issues review), Reports on Progress in Physics **79**, 032601 (2016).

[2] W. O. Hancock and J. Howard, Processivity of the Motor Protein Kinesin Requires Two Heads , Journal of Cell Biology **140**, 1395 (1998), https://rupress.org/jcb/article-pdf/140/6/1395/1490607/32983.pdf.

[3] J. O. Wirth, L. Scheiderer, T. Engelhardt, J. Engelhardt,

J. Matthias, and S. W. Hell, Minflux dissects the unimpeded walking of kinesin-1, Science **379**, 1004 (2023), https://www.science.org/doi/pdf/10.1126/science.ade2650.

[4] Y. Okada, H. Higuchi, and N. Hirokawa, Processivity of the single-headed kinesin kif1a through biased binding to tubulin, Nature **424**, 574–577 (2003).

[5] K. I. Schimert, B. G. Budaitis, D. N. Reinemann, M. J. Lang, and K. J. Verhey, Intracellular cargo transport by single-headed kinesin motors, Proceedings of the National Academy of Sciences **116**, 6152 (2019), https://www.pnas.org/doi/pdf/10.1073/pnas.1817924116.

[6] See supplemental material at http://link.aps.org/supplemental/, .

[7] F. Ruhnow, L. Klop, and S. Diez, Challenges in estimating the motility parameters of single processive motor proteins, Biophysical Journal **113**, 2433 (2017).

[8] C. Bustamante, D. Keller, and G. Oster, The physics of molecular motors, Accounts of chemical research **34**, 412 (2001).

[9] H. Wang and G. Oster, The stokes efficiency for molecular motors and its applications, Europhysics Letters **57**, 134 (2002).

[10] T. Schmiedl and U. Seifert, Efficiency of molecular motors at maximum power, Europhysics Letters **83**, 30005 (2008).

[11] S. Roychowdhury, G. Saraswat, S. Salapaka, and M. Salapaka, On control of transport in brownian ratchet mechanisms, Journal of Process Control **27**, 76 (2015).

[12] A. Yethiraj and A. van Blaaderen, A colloidal model system with an interaction tunable from hard sphere to soft and dipolar, nature **421**, 513 (2003).

[13] G. L. Hunter and E. R. Weeks, The physics of the colloidal glass transition, Reports on progress in physics **75**, 066501 (2012).

[14] U. Gasser, E. R. Weeks, A. Schofield, P. Pusey, and D. Weitz, Real-space imaging of nucleation and growth in colloidal crystallization, Science **292**, 258 (2001).

[15] A. Libál, D. Y. Lee, A. Ortiz-Ambriz, C. Reichhardt, C. J. Reichhardt, P. Tierno, and C. Nisoli, Ice rule fragility via topological charge transfer in artificial colloidal ice, Nature communications **9**, 4146 (2018).

[16] S. Erbas-Cakmak, D. A. Leigh, C. T. McTernan, and A. L. Nussbaumer, Artificial molecular machines, Chemical Reviews **115**, 10081 (2015), pMID: 26346838, https://doi.org/10.1021/acs.chemrev.5b00146.

[17] B. Lau, O. Kedem, J. Schwabacher, D. Kwasnieski, and E. A. Weiss, An introduction to ratchets in chemistry and biology, Mater. Horiz. **4**, 310 (2017).

[18] C. O. Reichhardt and C. Reichhardt, Ratchet effects in active matter systems, Annual Review of Condensed Matter Physics **8**, 51 (2017).

[19] S. G. Moorjani, L. Jia, T. N. Jackson, and W. O. Hancock, Lithographically patterned channels spatially segregate kinesin motor activity and effectively guide microtubule movements, Nano Letters **3**, 633 (2003), https://doi.org/10.1021/nl034001b.

[20] G. Patil, P. Mandal, and A. Ghosh, Using the thermal ratchet mechanism to achieve net motility in magnetic microswimmers, Phys. Rev. Lett. **129**, 198002 (2022).

[21] G. Camacho, A. Rodriguez-Barroso, O. Martinez-Cano, J. R. Morillas, P. Tierno, and J. de Vicente, Experimental realization of a colloidal ratchet effect in a non-newtonian fluid, Phys. Rev. Appl. **19**, L021001 (2023).

[22] J. S. Bader, R. W. Hammond, S. A. Henck, M. W. Deem, G. A. McDermott, J. M. Bustillo, J. W. Simpson, G. T. Mulhern, and J. M. Rothberg, Dna transport by a micromachined brownian ratchet device, Proceedings of the National Academy of Sciences **96**, 13165 (1999), https://www.pnas.org/doi/pdf/10.1073/pnas.96.23.13165.

[23] A. V. Arzola, M. Villasante-Barahona, K. Volke-Sepúlveda, P. Jákl, and P. Zemánek, Omnidirectional transport in fully reconfigurable two dimensional optical ratchets, Phys. Rev. Lett. **118**, 138002 (2017).

[24] J. Rousselet, L. Salome, A. Ajdari, and J. Prostt, Directional motion of brownian particles induced by a periodic asymmetric potential, Nature **370**, 446 (1994).

[25] L. P. Faucheux, L. Bourdieu, P. Kaplan, and A. J. Libchaber, Optical thermal ratchet, Physical review letters **74**, 1504 (1995).

[26] R. D. Astumian, Design principles for brownian molecular machines: how to swim in molasses and walk in a hurricane, Phys. Chem. Chem. Phys. **9**, 5067 (2007).

[27] R. D. Astumian and M. Bier, Fluctuation driven ratchets: Molecular motors, Phys. Rev. Lett. **72**, 1766 (1994).

[28] K. Svoboda, C. F. Schmidt, B. J. Schnapp, and S. M. Block, Direct observation of kinesin stepping by optical trapping interferometry, Nature **365**, 721 (1993).

[29] A. Yildiz, M. Tomishige, R. D. Vale, and P. R. Selvin, Kinesin walks hand-over-hand, Science **303**, 676 (2004).

[30] T. Deguchi, M. K. Iwanski, E.-M. Schentarra, C. Heidebrecht, L. Schmidt, J. Heck, T. Weihs, S. Schnorrenberg, P. Hoess, S. Liu, et al., Direct observation of motor protein stepping in living cells using minflux, Science **379**, 1010 (2023).

[31] S. Sudhakar, M. K. Abdosamadi, T. J. Jachowski, M. Bugiel, A. Jannasch, and E. Schäffer, Germanium nanospheres for ultraresolution picotensiometry of kinesin motors, Science **371**, eabd9944 (2021).

[32] R. D. Astumian, Thermodynamics and kinetics of a brownian motor, science **276**, 917 (1997).

[33] J. Howard and R. Clark, Mechanics of motor proteins and the cytoskeleton, Appl. Mech. Rev. **55**, B39 (2002).

[34] R. D. Vale, The molecular motor toolbox for intracellular transport, Cell **112**, 467 (2003).

[35] D. Oriola and J. Casademunt, Cooperative force generation of kif1a brownian motors, Phys. Rev. Lett. **111**, 048103 (2013).

[36] M. Hayakawa, Y. Kishino, and M. Takinoue, Collective ratchet transport generated by particle crowding under asymmetric sawtooth-shaped static potential, Advanced Intelligent Systems **2**, 2000031 (2020), https://onlinelibrary.wiley.com/doi/pdf/10.1002/aisy.202000031.

[37] F. J. Cao, L. Dinis, and J. M. R. Parrondo, Feedback control in a collective flashing ratchet, Phys. Rev. Lett. **93**, 040603 (2004).

[38] Y. Okada and N. Hirokawa, Mechanism of the single-headed processivity: Diffusional anchoring between the k-loop of kinesin and the c terminus of tubulin, Proceedings of the National Academy of Sciences **97**, 640 (2000), https://www.pnas.org/doi/pdf/10.1073/pnas.97.2.640.

[39] F. Martínez-Pedrero, F. Ortega, J. Codina, C. Calero, and R. G. Rubio, Controlled disassembly of colloidal aggregates confined at fluid interfaces using magnetic dipolar interactions, Journal of Colloid and Interface Science **560**, 388 (2020).

[40] M. W. Gramlich, L. Conway, W. H. Liang, J. A. Labastide, S. J. King, J. Xu, and J. L. Ross, Single molecule investigation of kinesin-1 motility using engi-

neered microtubule defects, Scientific Reports **7**, 44290 (2017).

[41] W. Zhang, Z. Zhang, S. Fu, Q. Ma, Y. Liu, and N. Zhang, Micro/nanomotor: A promising drug delivery system for cancer therapy, ChemPhysMater **2**, 114 (2023).

[42] J. Martín-Roca, F. Ortega, C. Valeriani, R. G. Rubio, and F. Martínez-Pedrero, Magnetic colloidal currents guided on self-assembled colloidal tracks, Advanced Functional Materials **33**, 2306541 (2023), https://onlinelibrary.wiley.com/doi/pdf/10.1002/adfm.202306541.